\documentclass[letter, journal]{IEEEtran}

\usepackage{cite}
\usepackage{amsmath,amssymb,amsfonts,bm,amsthm}
\usepackage{algorithm}
\usepackage{algorithmic}
\usepackage{graphicx}
\usepackage{subcaption}
\usepackage{textcomp}
\usepackage{xcolor}
\definecolor{r}{named}{red}
\usepackage{dsfont}
\usepackage{xspace,mathtools}
\usepackage{bm}
\usepackage{multicol}
\usepackage{float}
\usepackage{stfloats}
\usepackage{optidef}
\def\BibTeX{{\rm B\kern-.05em{\sc i\kern-.025em b}\kern-.08em
		T\kern-.1667em\lower.7ex\hbox{E}\kern-.125emX}}

\newcommand{\set}[1]{\ensuremath{\mathcal{#1}}\xspace} 
\newcommand{\mat}[1]{\ensuremath{\mathbf{#1}}\xspace} 
\renewcommand{\vec}[1]{\ensuremath{\mathbf{#1}}\xspace} 

\DeclarePairedDelimiter\abs{\lvert}{\rvert}
\DeclarePairedDelimiter\norm{\lVert}{\rVert}

\newcommand{\sinr}{\textsf{SINR}}

\newcommand{\rcs}{\textsf{rcs}}

\begin{document}

    \title{Load Balanced ISAC Systems for URLLC Users}
	\author{
  Shivani Singh, Amudheesan Nakkeeran, Prem Singh, Ekant Sharma, and Jyotsna Bapat
\thanks{Shivani Singh, Prem Singh, and Jyotsna Bapat are with the ECE department at IIIT Bangalore. Amudheesan Nakkeeran is with IIITB COMET Foundation and the ECE department at IIIT Bangalore. Ekant Sharma is with the ECE department at IIT Roorkee.} 
}
	
	
	\maketitle
    
    \begin{abstract}
%
This paper presents an energy‑efficient downlink cell‑free massive multiple‑input multiple‑output (CF‑mMIMO) integrated sensing and communication (ISAC) network that serves ultra‑reliable low‑latency communication (URLLC) users while simultaneously detecting a target. We propose a load‑balancing algorithm that minimizes the total network power consumption; including transmit power, fixed static power, and traffic‑dependent front haul power at the access points (APs) without degrading system performance. To this end, we formulate a mixed‑integer non‑convex optimization problem and introduce an iterative joint power allocation and AP load balancing (JPALB) algorithm. The algorithm aims to reduce total power usage while meeting both the communication quality‑of‑service (QoS) requirements of URLLC users and the sensing QoS needed for target detection. Proposed JPALB algorithm for ISAC systems was simulated with maximum‑ratio transmission (MRT) and regularized zero‑forcing (RZF) precoders. Simulation results show approximately $33\%$ reduction in power consumption, using JPALB algorithm compared to a baseline with no load balancing, without compromising communication and sensing QoS requirements.
    \end{abstract} 
    
    \section{Introduction}
    \bstctlcite{BSTcontrol2}
    Integrated sensing and communication (ISAC) is an integral component of the international mobile telecommunications (IMT)-2030 report for its application in industrial automation and vehicle-to-everything (V2X)~technology \cite{imt2030}. It allows harmonious coexistence of sensing and communication in a single system, improving spectrum efficiency and reduce hardware costs. The applications envisioned for the ISAC technology require ultra-reliable and low-latency communication (URLLC) to enable mission-critical functions, e.g. remote control of network-enabled industrial robots, autonomous driving~\cite{Behdad2024Interplay}.

    Cell-Free massive multiple-input multiple-output (CF-mMIMO) technology has been widely researched in recent times as a viable architecture to realize multi-static ISAC networks \cite{Behdad2024Multi-Static, Palhano2025Power, Elfiatour2023Cell-Free, Elfiatour2025Multiple-Target, Demirhan2024Cell-Free, Xia2025Joint, Nguyen2024Multi-Static, Behdad2024Interplay, Nikbakht2025MIMO,Singh2025Target}. Reference \cite{Behdad2024Multi-Static} developed a maximum aposteriori ratio test detector for detecting the presence of a target at a known location and proposed a downlink power allocation algorithm to maximize the sensing signal-to-interference-plus-noise-ratio (SINR) in a CF-mMIMO ISAC system while meeting communication requirements. In \cite{Palhano2025Power}, the authors studied an uplink sensing scenario in a CF-mMIMO ISAC network where the user equipment (UEs) adopt a multibeam scheme for communication and sensing symbols, designed using singular value decomposition (SVD) beamforming and angular beamforming. The authors in \cite{Elfiatour2023Cell-Free} formulated and solved a max-min fairness problem where the minimum spectral efficiency (SE) of UEs is maximized subject to per-access point (AP) power budget and mainlobe-to-average-sidelobe ratio (MASR) constraints by power allocation and AP mode selection (communication or sensing) in a CF-mMIMO ISAC network. The authors in \cite{Elfiatour2025Multiple-Target} extended the work to consider multiple targets and partial zero-forcing (PZF) precoders. Reference \cite{Demirhan2024Cell-Free} proposed various beamforming approaches to maximize the sensing SINR in a CF-mMIMO ISAC network while ensuring a minimum required communication SINR. In \cite{Xia2025Joint}, the authors performed joint location and velocity estimation of a target in a CF-mMIMO ISAC system and derive relevant Cramer-Rao lower bounds (CRLB), while the authors of \cite{Nguyen2024Multi-Static} derive closed form expressions for achievable communication rate and CRLB for a target's angle estimation. 

    While the aforementioned works consider an infinite blocklength regime\footnote{Shannon capacity requires infinite blocklength codewords for perfect reliability in the non-asymptotic regime, which is not appropriate for achievable data rate under URLLC constraints like transmission latency and finite packet length \cite{Zhang2019Statistical}.}, indicating a focus on enhanced mobile broadband (eMBB) scenario, a few recent works have also investigated the finite blocklength regime suitable serving for URLLC users using CF-mMIMO ISAC networks. Reference \cite{Behdad2024Interplay} proposes a successive convex approximation (SCA) based power allocation algorithm to maximize the energy efficiency of a CF-mMIMO ISAC network while ensuring URLLC reliability and sensing SINR requirements. However, they have not considered energy-efficient load balancing to reduce power consumption outside the peak traffic hours. The authors in \cite{Nikbakht2025MIMO} considered a bistatic MIMO ISAC network where URLLC transmissions are triggered when the presence of a target is detected through copies of eMBB transmissions reflected by the target. 

    Since future networks are expected to support almost a trillion devices \cite{imt2030}, network power consumption has become an important design factor \cite{Chien2020Joint, Demir2024Cell-Free}. In \cite{Chien2020Joint}, the authors proposed low-complexity algorithms to decide which APs to turn off in a CF-mMIMO network while guaranteeing SE requirements. Reference \cite{Demir2024Cell-Free} considered a CF-mMIMO network running on an open radio access network (ORAN) architecture and performed end-to-end resource orchestration, including turning off a subset of APs while maximizing sum SE. Note that \cite{Chien2020Joint, Demir2024Cell-Free} did not consider combination of ISAC and URLLC UEs, which is required for applications such as autonomous driving that harness the benefits of both ISAC and URLLC. It is therefore imperative to develop solutions that reduce the network power consumption while still ensuring the required communication and sensing requirements for CF-mMIMO ISAC system in the existance of URLLC UEs. In this context, we summarize the contributions of our paper now. 
    \begin{itemize}
        \item We introduce a model that considers minimization of power consumption by turning communication APs off (load balancing) in cell-free massive MIMO ISAC systems, simultaneously serve URLLC users and detecting target. The power consumption model includes the transmit power, fixed static power, and traffic-dependent power consumed by the APs in a CF-mMIMO ISAC network.
        \item Forming the load balancing problem as a mixed integer non-convex optimization problem, we design an iterative algorithm based on the difference of convex functions and convex-concave methods to jointly perform power allocation and communication APs load balancing while meeting the URLLC rate and sensing SINR constraints. 
        \item We illustrate the efficacy of our algorithm using simulations by comparing it to a baseline case which performs only power allocation without any load balancing, demonstrating that our algorithm leads to $\sim33\%$ reduction in total average power consumed in the network.  
    \end{itemize}
    
    \section{System Model}
	We consider a CF mMIMO multi-static ISAC system, wherein a set $\mathcal{K}$ of $K$ communication access points (APs) collaborate via a central processing unit (CPU) to serve a set $\mathcal{U}$ of $U$ single-antenna ultra-reliable low-latency communications (URLLC) user equipment (UEs), by leveraging the cloud radio access network (C-RAN) architecture. Additionally, a set $\mathcal{R}$ of $R$ APs act as sensing (or receive) APs to detect the presence of a target at a specific location by using the downlink communication signals reflected off the target. All the $K + R$ APs are equipped with $M$ antennas each and are connected via fronthaul links to the CPU which provides perfect synchronization among the APs. We assume time division duplex (TDD) protocol and block-fading channel model, in which each coherence block contains $\tau$ symbols, out of which $\tau_d$ symbols are used for downlink data transmission and the rest for pilot transmission. During each coherence time interval, the receive APs collect the reflection of the $\tau$ symbols off the target for sensing. Since URLLC traffic is latency-sensitive, we consider the case where the transmitted packets (codewords) span $\tau_d$ symbols and can thus be transmitted in one coherence period \cite{Behdad2024Interplay}. 

    \subsection{Communication model}
    We consider downlink data transmission from a load balancing perspective where, at any given time instant, only a subset $\mathcal{A} \subseteq \mathcal{K}$ of communication APs can actively serve the UEs while the remaining communication APs in the set $\mathcal{K} \backslash \mathcal{A}$ can enter sleep mode to reduce power consumption. Note that the decision of which communication APs should be included in $\mathcal{A}$ impacts not just the communication quality of service (QoS) of the UEs, but also the accuracy of target detection by sensing APs since only a subset of communication APs are active in the downlink. Furthermore, the location of these active APs influences both the communication $\sinr$ and sensing $\sinr$, and hence the performance of the system as a whole. The CPU decides the form-factor of $\mathcal{A}$, and we capture this through binary indicators $\{\alpha_k\}_{k \in \mathcal{K}}$ as
	\begin{equation} \label{eq_active_AP_indicator}
		\alpha_k = \begin{cases}
			1, \quad k \in \mathcal{A} \subseteq \mathcal{K}, \\
			0, \quad \text{otherwise.}
		\end{cases}
	\end{equation}
    The communication AP activity (active/sleep) indicators $\{\alpha_k\}_{k \in \mathcal{K}}$ can be communicated to the APs via fronthaul interfaces. The CPU also designs, for each coherence time interval, AP-specific unit-norm precoding vectors $\left\lbrace \vec{w}_{i,k} \in \mathds{C}^{M \times 1} \right\rbrace_{i \in \set{U}, k \in \set{A}}$ for AP $k$ and UE $i$, and UE-specific power coefficients $\left\lbrace \sqrt{\rho_{i,k}} \right\rbrace_{i \in \mathcal{U}, k \in \set{K}}$ for AP $k$ and UE $i$. The transmitted signal from AP $k \in \set{K}$ at time instance $l$ is
    \begin{equation} \label{eq_tx_signal}
        \vec{x}_k[l] = \sum_{i \in \mathcal{U}} \vec{w}_{i,k} \alpha_k \sqrt{\rho_{i,k}} s_i[l] = \alpha_k\mat {W}_{k}\mat{D}_s[l]\boldsymbol{\rho}_k',
    \end{equation}
    where $s_i[l]$ is a zero-mean unit-power communication symbol intended for UE $i$. In \eqref{eq_tx_signal}, $\mat {W}_{k} = [\vec{w}_{1,k}, \dots, \vec{w}_{U,k}]  \in \mathds{C}^{M \times U}$ is the precoding matrix, $\mat{D}_s = \text{diag}\{s_1, \dots, s_U \}  \in \mathds{C}^{U \times U}$ is the data matrix and $\boldsymbol{\rho}_{k}' = \left[ \sqrt{\rho_{1,k}}, \sqrt{\rho_{2,k}}, \dots, \sqrt{\rho_{U,k}} \right]^{\text{T}}  \in \mathds{C}^{U \times 1}$ is the power coefficient vector. 
   \subsubsection{Power consumption}
    From \eqref{eq_tx_signal}, we calculate the average transmit power of AP $k \in \set{K}$, when it is active, as
    \begin{equation} \label{eq_tx_power_full}
        P_k = \mathds{E} \left[ \norm{\vec{x}_k[l]}^2 \right] = \alpha_k \sum_{i \in \set{U}} \rho_{i,k} \mathds{E} \left[ \norm{\vec{w}_{i,k}}^2 \right],
    \end{equation}
    which is a function of $\alpha_k$ and $\{\rho_{i,k}\}_{i \in \set{U}}$, with each AP having a power budget of $P_{\text{max}}$ Watts. The overall power consumption of the network is then given by \cite{Chien2020Joint}
    \begin{align}
        P_{\text{total}} \left( \{\alpha_k\}, \{\rho_{i,k}\} \right) = \eta\sum_{k \in \set{K}} \sum_{i \in \set{U}} \alpha_k \rho_{i,k} \mathds{E} \norm{\vec{w}_{i,k}}^2 + \nonumber \\ \sum_{k \in \set{K}} \alpha_k P_0 + B\sum_{k \in \set{K}} \sum_{i \in \set{U}} \alpha_k P_k^{\text{tra}} R_i,
    \end{align}
    where $\eta$ is the power amplifier efficiency, $P_0$ is the static power consumption of an AP even when it is not transmitting any signal, $P_k^{\text{tra}}$ is the traffic-dependent fronthaul power consumed by an AP, $B$ is the system bandwidth, and $R_i$ is the communication rate of UE $i$. Here, the static power consumption at each AP is modeled as $P_{0} = P_{\text{fixed}}^{\text{fh}} + P_{\text{hw}} M$ \cite{Chien2020Joint},
    where $P_{\text{fixed}}^{\text{fh}}$ is the traffic-independent fixed power consumed by each fronthaul link, and $P_{\text{hw}}$ is the hardware power consumption by each antenna. 
	
   We denote the channel between communication AP $k$ and UE $i$ by $\vec{h}_{i,k} \in \mathds{C}^{M \times 1}$, which is modeled with Rayleigh fading as $\vec{h}_{i,k} \sim \mathcal{N}_{\mathds{C}} \left( \vec{0}, \beta_{i,k} \mat{I}_{M} \right)$ with $\beta_{i,k}$ being the corresponding large-scale fading coefficient. The signal received at UE $i$ after combining with the channel $\vec{h}_{i,k}$  can be written as
	\begin{align} \label{eq_ue_rx_signal}
		y_i[l] &= \sum_{k \in \mathcal{K}} \sum_{i \in \mathcal{U}} \vec{h}_{i,k}^{\text{H}}\vec{w}_{i,k} \alpha_k \sqrt{\rho_{i,k}} s_i[l] + n_i[l], \nonumber \\
        &= \underbrace{\sum_{k \in \mathcal{K}} \alpha_k \sqrt{\rho_{i,k}} \vec{h}_{i,k}^{\text{H}}\vec{w}_{i,k} s_i[l] }_{\text{Desired signal (DS}_{i})} \nonumber \\
		& ~~~+ \sum_{j \in \mathcal{U} \backslash i} \underbrace{ \sum_{k \in \mathcal{K}} \alpha_k \sqrt{\rho_{j,k}} \vec{h}_{i,k}^{\text{H}}\vec{w}_{j,k} s_j[l] }_{\text{Interference due to UE $j$ (IUI}_{i,j})} ~+~ n_i[l],
	\end{align}
	where $n_i[l] \sim \mathcal{N}_{\mathds{C}}(0,\sigma_{n}^2)$ is the channel noise. 
	Using \eqref{eq_ue_rx_signal}, the communication SINR seen by UE $i$ can then be expressed as
	\begin{equation} \label{UE_SINR}
		\gamma_i = \frac{\abs{\text{DS}_i}^2}{ \sum_{j \in \mathcal{U} \backslash i} \abs{\text{IUI}_{i,j}}^2 + \sigma_{n}^2 } = \frac{\abs{\vec{b}_i^{\text{T}} \boldsymbol{\rho}_i}^2}{ \sum_{j \in \set{U}} \boldsymbol{\rho}_j^{\text{T}} \mat{C}_{i,j} \boldsymbol{\rho}_{j} + \sigma_n^2 },
	\end{equation}
	where $\boldsymbol{\rho}_i = \left[\alpha_1 \sqrt{\rho_{i,1}}, \alpha_2\sqrt{\rho_{i,2}}, \dots, \alpha_K\sqrt{\rho_{i,K}} \right]^{\text{T}}$, $\vec{b}_i = \left[ \mathds{E}\left[ \vec{h}_{i,1}^{\text{H}} \vec{w}_{i,1} \right], \mathds{E}\left[ \vec{h}_{i,2}^{\text{H}} \vec{w}_{i,2} \right], \dots, \mathds{E}\left[ \vec{h}_{i,K}^{\text{H}} \vec{w}_{i,K} \right] \right]^{\text{T}}$, and
    \begin{align}
        &\left[\mat{C}_{i,j}\right]_{x,y} = \begin{cases}
            \mathds{E}\left[ \vec{h}_{i,x}^{\text{H}} \vec{w}_{j,x} \vec{w}_{j,y}^{\text{H}} \vec{h}_{i,y}^{*} \right], \quad i \neq j, \\
            \mathds{E}\left[ \vec{h}_{i,x}^{\text{H}} \vec{w}_{i,x} \vec{w}_{i,y}^{\text{H}} \vec{h}_{i,y}^{*} \right] - \left[ \vec{b}_i \right]_x \left[ \vec{b}_i \right]_x^{*},~~ i = j.
        \end{cases}
    \end{align}
     Here $\left[\mat{C}_{i,j}\right]_{x,y}$ is the $(x,y)$-th entry of $\mat{C}_{i,j}$, and $\left[ \vec{b}_i \right]_x$ is the $x$-th entry of $\vec{b}_i$. Using \eqref{UE_SINR}, the approximate achievable downlink data rate for the $i$-th URLLC UE in bits/sec/Hz is \cite{Peng2023Resource}
	\begin{equation} \label{eq_urllc_rate_full}
	    R_i \approx \frac{\tau_d}{\tau \ln(2)} \ln(1 + \gamma_i) - \frac{Q^{-1} \left( \epsilon_i \right)}{\tau\ln(2)} \sqrt{\tau_d V_i},
	\end{equation} 
	where $\epsilon_i$ is the required decoding error probability, $Q^{-1}(\cdot)$ is the inverse of the Gaussian-Q function, and $V_i = 1 - \left( 1 + \gamma_i \right)^{-1}$ is the channel dispersion. 

	\subsection{Sensing Model} 
	Recall that the system has a set $\mathcal{R}$ of $R$ sensing APs that are involved in multi-static target detection by collecting all communication signal from all active communication APs and their reflections through the target (if present). Taking advantage of the C-RAN architecture, the sensing APs can remove the known target-free components of the received signal. The resulting received signal $\vec{y}_r \in \mathds{C}^{M \times 1}$ at sensing AP $r$ in time instance $l$ is given by
	\begin{align} \label{eq_rxAP_signal}
	    \vec{y}_r [l] = \sum_{k \in \set{K}} &\underbrace{\omega_{r,k} \vec{a}\left( \phi_r, \theta_r \right) \vec{a}^{\text{T}}\left( \varphi_k, \vartheta_k \right) \vec{x}_k[l]}_{\text{Reflections off target}} \nonumber \\
        & + \sum_{k \in \set{K}} \underbrace{\mat{H}_{r,k} \vec{x}_k[l]}_{\text{Undesired reflections by clutter}} + \vec{n}_r[l],
	\end{align}
    where $\vec{n}_r[l] \sim \mathcal{N}_{\mathds{C} \left( \vec{0}, \sigma_n^2 \mat{I}_M \right)}$ is the noise at sensing AP $r$ and $\omega_{r,k} \sim \mathcal{N}_{\mathbb{C}} \left(0, \sigma_{\rcs}^2\right)$ is the radar-cross section (RCS) of the target that includes channel gain of the path through the target between communication AP $k$ and sensing AP $r$ and follows Swerling-I model \cite{Behdad2024Interplay}. The matrix $\mat{H}_{r,k} \in \mathds{C}^{M \times M}$ models environmental clutter between communication AP $k$ and sensing AP $r$.

	\section{Problem Formulation and Proposed Solution} 
    The CPU collects all the $\left\lbrace \vec{y}_r[l] \right\rbrace_{r \in \set{R}}$ from the sensing APs across all the $\tau_d$ communication symbols in the sensing duration and calculates the sensing SINR as 
    \begin{equation*} \label{eq_sens_SINR}
        \sinr_s = \frac{ \sigma_{\text{rcs}}^2 \sum_{l=1}^{\tau_d} \sum_{r \in \set{R}} \sum_{k \in \set{K}} \left[ \left(\boldsymbol{\rho}_{k}'\right)^{\text{T}} \mat{A}_{r,k}[l] \boldsymbol{\rho}_{k}' \right] }{\sum_{l=1}^{\tau_d} \sum_{r \in \set{R}} \sum_{k \in \set{K}} \left[ \left(\boldsymbol{\rho}_{k}'\right)^{\text{T}} \mat{B}_{r,k}[l] \boldsymbol{\rho}_{k}' \right] + \tau_{d}MR\sigma_n^2},
    \end{equation*}
    where the positive semidefinite matrices $\mat{A}_{r,k}[l]$ and $\mat{B}_{r,k}[l], ~\forall r \in \mathcal{R}, k \in \mathcal{K}, l = 1,\dots,\tau_d$ are computed similar to \cite{Behdad2024Multi-Static} as
    \begin{align} \label{eq_matA}
        \mat{A}_{r,k}[l] = \mat{D}_s^{\text{H}}[l] \mat{W}_k^{\text{H}} & \vec{a}^{*}\left( \varphi_k, \vartheta_k \right) \vec{a}^{\text{H}}\left( \phi_r, \theta_r \right) \nonumber \\
        &\times \vec{a}\left( \phi_r, \theta_r \right) \vec{a}^{\text{T}}\left( \varphi_k, \vartheta_k \right) \mat{W}_k \mat{D}_s[l], 
    \end{align}
    \begin{equation} \label{eq_matB}
        \mat{B}_{r,k}[l] = \mat{D}_s^{\text{H}}[l] \text{tr} \left( \mat{R}_{\text{rx},(r,k)} \right) \mat{W}_k^{\text{H}} \mat{R}_{\text{tx},(k,r)}^{\text{T}} \mat{W}_k \mat{D}_s[l].
    \end{equation}
    Here $\mat{R}_{\text{rx},(r,k)}$ is the spatial correlation matrix associated with sensing AP $r$ in the direction of communication AP $k$, and $\mat{R}_{\text{tx},(k,r)}$ is the spatial correlation matrix associated with communication AP $k$ in the direction of sensing AP $r$.
    
	We aim to minimize the total power consumed in the cell-free massive MIMO ISAC network by choosing the power allocation vectors $\{\boldsymbol{\rho}\}$ and communication AP activity indicators  $\{\alpha_k\}_{k \in \mathcal{K}}$ such that constraints on URLLC downlink data rate, sensing SINR, and per-communication-AP power budget are satisfied. This optimization problem, with variables $\boldsymbol{\rho}\geq \vec{0}$ and $\{\alpha_k\}_{k \in \mathcal{K}}$, is formulated as follows 
	\begin{mini!}{}{ P_{\text{total}}\left( \{\alpha_k\}_{k \in \mathcal{K}}, \{\rho_{i,k}\}_{i \in \mathcal{U}, k \in \set{K}} \right) }{\label{opti_original_problem}}{\label{opti_original_objective}}
		\addConstraint{R_i \geq R_{\text{min}}, ~\forall i \in \mathcal{U}} \label{opti_qos_constraint}
		\addConstraint{\sinr_{\text{s}} \geq \Gamma_{\text{th}}^{\text{sen}}} \label{opti_sensing_constraint}
		\addConstraint{P_k \leq P_{\text{max}}, \quad \forall k \in \mathcal{A}} \label{opti_power_constraint}
		\addConstraint{\alpha_k \in \{0,1\}, \quad \forall k \in \mathcal{K}.} \label{opti_bin_constraint}
	\end{mini!}
	The optimization problem in \eqref{opti_original_problem} is a mixed-integer non-convex problem (MINCP), where \eqref{opti_qos_constraint} ensures that the URLLC data rate exceeds a minimum threshold $R_{\text{min}}$, \eqref{opti_sensing_constraint} is the sensing SINR constraint that ensures target detection, \eqref{opti_power_constraint} is the per-communication-AP power constraint, and \eqref{opti_bin_constraint} restricts the communication AP activity indicators to be either 0 (turned off) or 1 (active). Note that \eqref{opti_qos_constraint}, through \eqref{eq_urllc_rate_full}, ensures that the required reliability is met. Since the packets are assumed to be transmitted within one coherence period, the latency requirements are already met.
    \subsection{ Proposed Solution}
    To solve the problem in \eqref{opti_original_problem}, we first deal with the constraints, followed by designing an iterative joint power allocation and AP load balancing (JPALB) algorithm based on convex-concave and difference of convex functions methods. 
    \subsubsection{URLLC data rate constraints}
    To overcome the difficulty in dealing with the non-convex expression for the URLLC rate in \eqref{eq_urllc_rate_full}, since $V_i < 1$, we first lower bound the data rate as 
    \begin{equation} \label{eq_urllc_rate_lb}
        R_i ~\geq~ \frac{\tau_d}{\tau \ln(2)} \ln(1 + \gamma_i) - \frac{\sqrt{\tau_d}Q^{-1} \left( \epsilon_i \right)}{\tau\ln(2)} \triangleq R_{i}^{\text{lb}}.
    \end{equation}    
     Then, we constrain $R_i^{\text{lb}} \geq R_{\text{min}}$ instead of constraining the actual rate as in \eqref{opti_qos_constraint}. Using \eqref{eq_urllc_rate_lb}, we arrive at the following equivalent constraint on the SINR at UE $i$
    \begin{align} \label{eq_urllc_rate_constraint}
        \gamma_i \geq \exp \left\lbrace \frac{1}{\sqrt{\tau_d}} \left( \frac{\tau \ln(2)}{\sqrt{\tau_d}} R_{\text{min}} + Q^{-1}(\epsilon_i) \right) \right\rbrace - 1 \triangleq \gamma_{\text{th}}.
    \end{align}
    We then rewrite \eqref{eq_urllc_rate_constraint}, using \eqref{UE_SINR}, as a convex second-order cone (SOC) constraint, similar to \cite{Demir2024Cell-Free}, as 
\begin{equation} \label{eq_sinr_constraint}
    \norm*{\left[ \mat{C}_{i,1}^{\frac{1}{2}} \boldsymbol{\rho}_1,  \mat{C}_{i,2}^{\frac{1}{2}} \boldsymbol{\rho}_2, \dots \mat{C}_{i,U}^{\frac{1}{2}} \boldsymbol{\rho}_U, \sigma_n\right]} \leq \sqrt{\frac{\abs{\vec{b}_i^{\text{T}} \boldsymbol{\rho}_i}^2}{\gamma_{\text{th}}}}, \quad \forall i \in \set{U}.
\end{equation}

    \subsubsection{Sensing SINR constraints}
    We first express the sensing SINR constraints in \eqref{opti_sensing_constraint} as
    \begin{equation}
        \boldsymbol{\rho}^{\text{T}} \left( \Gamma_{\text{th}}^{\text{sen}} \mat{B} - \mat{A} \right) \boldsymbol{\rho} \leq -\Gamma_{\text{th}}^{\text{sen}} \tau_d M R \sigma_n^2,
    \end{equation}
    where $\boldsymbol{\rho} = \left[ \left(\boldsymbol{\rho}_{1}'\right)^{\text{T}}, \left(\boldsymbol{\rho}_{2}'\right)^{\text{T}}, \dots, \left(\boldsymbol{\rho}_{K}'\right)^{\text{T}} \right]^{\text{T}}$, and $\mat{A} = \text{blkdiag}\left\lbrace \mat{A}_1, \mat{A}_2, \dots, \mat{A}_K \right\rbrace$, $\mat{B} = \text{blkdiag}\left\lbrace \mat{B}_1, \mat{B}_2, \dots, \mat{B}_K \right\rbrace$ with $\mat{A}_k = \sum_{l=1}^{\tau_d} \sum_{r \in \set{R}} \mat{A}_{r,k}[l]$ and $\mat{B}_k = \sum_{l=1}^{\tau_d} \sum_{r \in \set{R}} \mat{B}_{r,k}[l]$. We then use a linear restriction around a feasible point $\boldsymbol{\rho}^{(c)}$ to arrive at the convex constraint
    \begin{align} \label{eq_sensing_sinr_constraints_final}
        \Gamma_{\text{th}}^{\text{sen}} \boldsymbol{\rho}^{\text{T}} \mat{B} \boldsymbol{\rho} &+ 2 \mathfrak{R} \left\lbrace \left( \boldsymbol{\rho}^{(c)} \right)^{\text{T}} (-\mat{A}) \boldsymbol{\rho} \right\rbrace \nonumber \\
        &\leq  -\Gamma_{\text{th}}^{\text{sen}} \tau_d M R \sigma_n^2 + \left( \boldsymbol{\rho}^{(c)} \right)^{\text{T}} (-\mat{A}) \boldsymbol{\rho}^{(c)}.
     \end{align}

    \subsubsection{Per-communication-AP power budget constraints}
    We rewrite the per-communication-AP power budget constraints in \eqref{opti_power_constraint} as convex SOC constraints, similar to \cite{Behdad2024Multi-Static}, as 
    \begin{equation} \label{eq_power_constraints}
        \norm{\mat{G}_k^{\text{T}} \boldsymbol{\rho}} \leq \alpha_k \sqrt{P_{\text{max}}},
    \end{equation}
    where $\mat{G}_k = \text{diag} \left\lbrace \norm{\vec{w}_{1,k}}_{\text{F}}, \norm{\vec{w}_{2,k}}_{\text{F}}, \dots, \norm{\vec{w}_{U,k}}_{\text{F}} \right\rbrace$.

    \subsubsection{AP activity indicator (binary) constraints}
    To address the non-convexity of the binary constraints in \eqref{opti_bin_constraint}, we first rewrite it as $\alpha_k - \alpha_k^2[m] \leq 0$ with $0 \leq \alpha_k \leq 1$, and then use the second-order Taylor approximation for $\alpha_k^2$ around a feasible point $\alpha_k^{(c)}$ to arrive at, $\forall k \in \set{K}$, 
    \begin{equation} \label{eq_bin_constraint_range_final}
    	0 \leq \alpha_k \leq 1, 
    \end{equation} 
    \begin{equation} \label{eq_bin_constraint_final}
    	\alpha_k - \left[ \left( \alpha_k^{(c)} \right)^2 + 2 \alpha_k^{(c)} \left( \alpha_k - \alpha_k^{(c)} \right) \right] \leq 0.
    \end{equation} 
    
    
    \subsubsection{Convergence issues due to the relaxed binary constraints} 
    To ensure that the $\{\alpha_k\}_{k \in \mathcal{K}}$ all converge to either 0 or 1, we add the left-hand side of \eqref{eq_bin_constraint_final} to the objective function in \eqref{opti_original_objective} to make it a difference of convex functions. This allows us to use the convex-concave procedure which iteratively solves a sequence of convex optimization problems to arrive at the solution. The solutions obtained in iteration $c$ can be expressed via the previous iterates as
    \begin{align} \label{eq_opt_form_final}
        &\boldsymbol{\rho}^{(c)}, \left\lbrace \alpha_k^{(c)} \right\rbrace_{k \in \set{K}} = \arg \min_{\boldsymbol{\rho}, \left\lbrace \alpha_k \right\rbrace_{k \in \set{K}}} \Bigg\lbrace P_{\text{total}} + \Bigg. \nonumber \\ 
        &\left. \sum_{k \in \set{K}} \left( \alpha_k - \left[ \left( \alpha_k^{(c-1)} \right)^2 + 2 \alpha_k^{(c-1)} \left( \alpha_k - \alpha_k^{(c-1)} \right) \right] \right) \right\rbrace.
    \end{align}
    subject to the constraints in \eqref{eq_sinr_constraint}, \eqref{eq_sensing_sinr_constraints_final}, \eqref{eq_power_constraints}, \eqref{eq_bin_constraint_range_final}, \eqref{eq_bin_constraint_final}.\\

        \begin{algorithm}
	\caption{Proposed Joint Power Allocation and AP Load Balancing (JPALB) Algorithm}\label{algo}
	\begin{algorithmic} [1]
		\STATE \textbf{Initialize:} 
		\STATE \hspace{\algorithmicindent} Set arbitrary feasible points for $\boldsymbol{\rho}^{(0)}$, and $\left\lbrace\alpha_k^{(0)}\right\rbrace_{k \in \mathcal{K}}$.
		\STATE \hspace{\algorithmicindent} Set iteration counter $c = 0$.
		\STATE \hspace{\algorithmicindent} Set tolerance for stopping condition $\eta = 10^{-3}$, and initial value of objective function $f^{(0)}$.
		\STATE \textbf{While} $c < c_{\text{m}}$ or $\abs{f^{(c)} - f^{(c-1)}} > \eta$ \textbf{do}:
		\STATE \hspace{\algorithmicindent} $~c \leftarrow c + 1$.
		\STATE \hspace{\algorithmicindent} Obtain $\boldsymbol{\rho}^{(c)}$, and $\left\lbrace\alpha_k^{(c)}\right\rbrace_{k \in \mathcal{K}}$ as the solution of \eqref{eq_opt_form_final} subject to the constraints in \eqref{eq_sinr_constraint}, \eqref{eq_sensing_sinr_constraints_final}, \eqref{eq_power_constraints}, \eqref{eq_bin_constraint_range_final}, \eqref{eq_bin_constraint_final}.
		\STATE \textbf{End}
        \STATE \textbf{Output:} Power allocation vectors $\boldsymbol{\rho}^{*}$, and communication AP activity indicators $\left\lbrace\alpha_k^{*}\right\rbrace_{k \in \mathcal{K}}$.
	\end{algorithmic}
    \end{algorithm}
    
    \subsubsection{Iterative JPALB Algorithm}
    We now present the proposed iterative JPALB scheme in Algorithm \ref{algo}, to solve the optimization problem in \eqref{opti_original_problem}. The loop between lines $5$ and $8$ in the $c$-th iteration generates $\boldsymbol{\rho}^{(c)}$ and $\left\lbrace \alpha_k^{(c)} \right\rbrace_{k \in \set{K}}$ by solving the optimization problem in \eqref{eq_opt_form_final} in each iteration, subject to the constraints in \eqref{eq_sinr_constraint}, \eqref{eq_sensing_sinr_constraints_final}, \eqref{eq_power_constraints}, \eqref{eq_bin_constraint_range_final}, \eqref{eq_bin_constraint_final}. The algorithm outputs the power allocation vectors $\boldsymbol{\rho}^{*}$ and the communication AP activity indicators $\left\lbrace\alpha_k^{*}\right\rbrace_{k \in \mathcal{K}}$ when either the absolute difference between the values of the objective function in successive iterations is within a tolerable limit, or if the number of iterations exceeds a predefined maximum.

    \begin{figure*}
     \centering
     \begin{subfigure}[b]{0.32\textwidth}
         \centering
         \includegraphics[width=\textwidth]{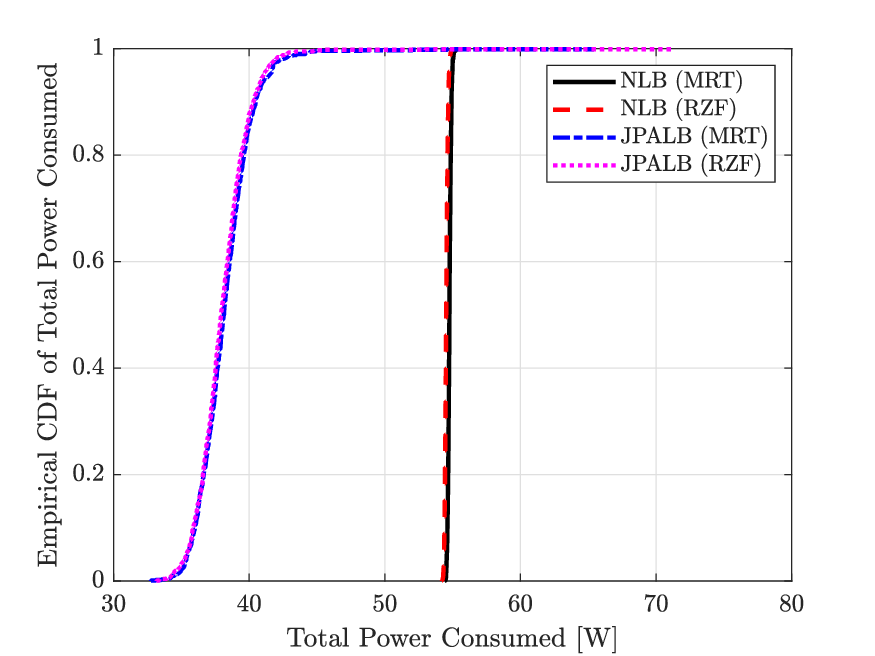}
         \caption{}
         \label{power-vs-sinr}
     \end{subfigure}
     \begin{subfigure}[b]{0.32\textwidth}
         \centering
         \includegraphics[width=\textwidth]{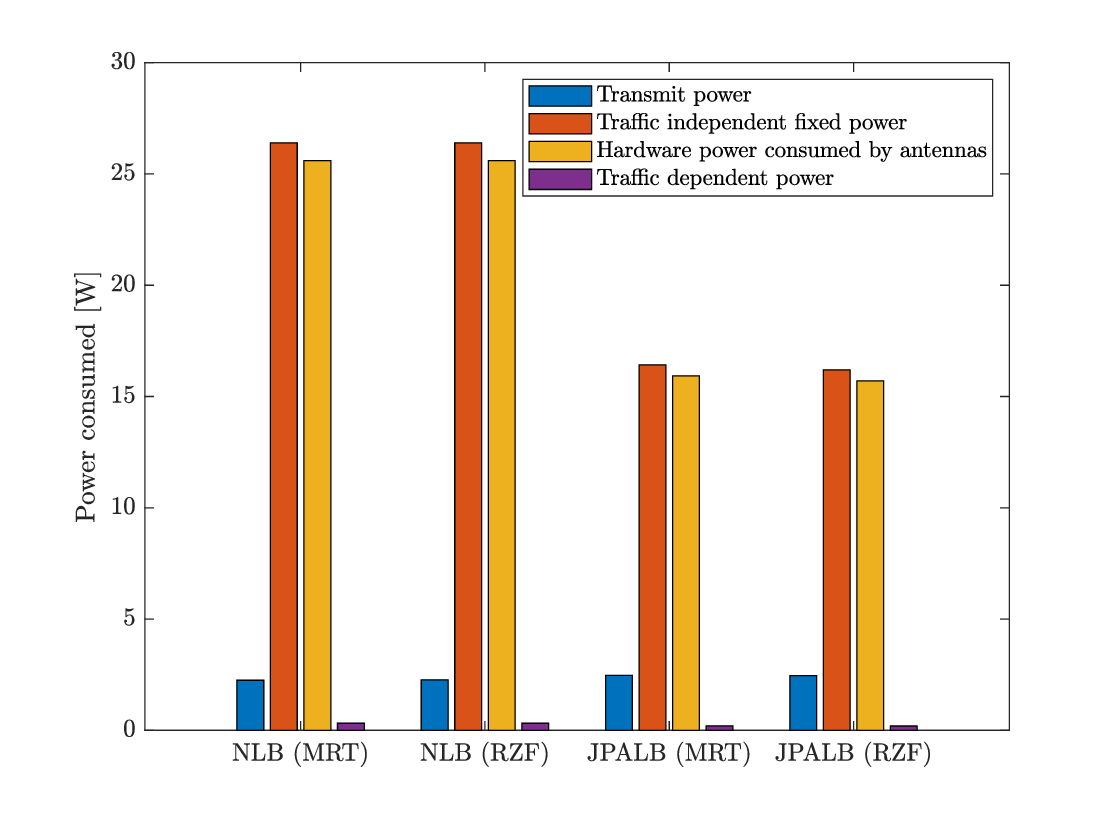}
         \caption{}
         \label{bar-plots}
     \end{subfigure}
     \begin{subfigure}[b]{0.32\textwidth}
         \centering
         \includegraphics[width=\textwidth]{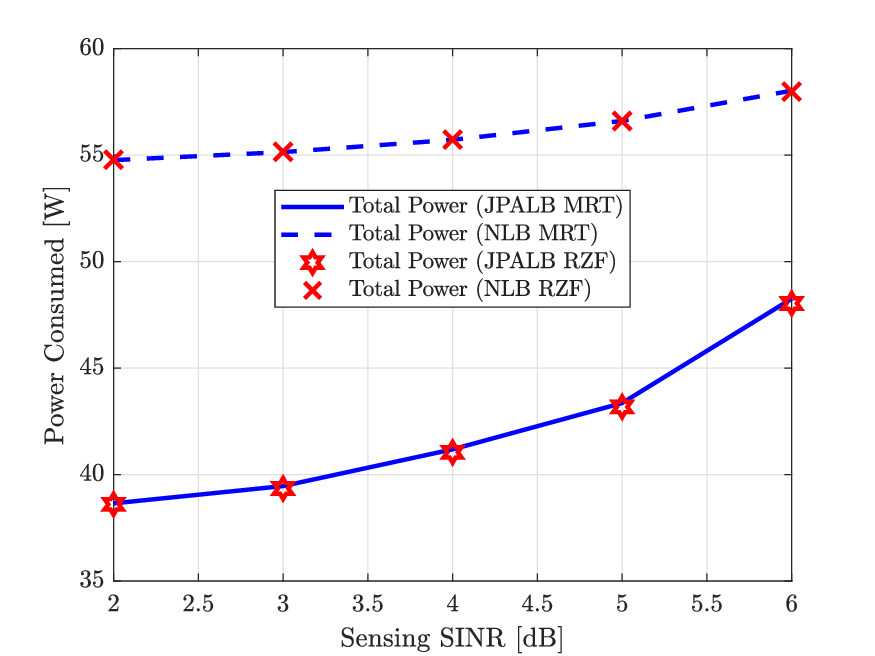} 
         \caption{}
         \label{power-vs-rcs}
     \end{subfigure}
     \hfill
    \caption{\small (a) CDF of average total power consumed [W]; (b) power consumed by various components in the network; (c) total power consumed for varying sensing $\sinr$ $\left(\Gamma_{\text{th}}^{\text{sen}}\right)$. For all plots, $K=32, ~R=2, ~U=8, ~M=4, ~R_{\text{min}} = 2 \text{ b/s/Hz }, ~\Gamma_{\text{th}}^{\text{sen}} = 2 \text{ dB}, ~\sigma_\text{rcs}^2 = 2 \text{ dBsm}$.}
        \label{fig:four-graphs}
\end{figure*}



    \section{Numerical Results and Discussion}

    We simulate the results for a cell-free MIMO ISAC setup in which AP locations are generated uniformly at random in a $500$ m $\times$ $500$ m area. The $R$ closest APs to the target, which appears at the center of the area, are designated as the sensing APs. We consider $K = 32$ communication APs, $R = 2$ sensing APs, and $U = 8$ UEs unless stated otherwise. Each AP has $M = 4$ antennas and maximum transmit power $P_\textsf{max} = 1$ W. The carrier frequency, bandwidth, and noise power are chosen to be $1.9$ GHz, $20$ MHz and $-94$ dBm respectively. Unless stated otherwise, the minimum required URLLC rate is $R_{\text{min}} = 1$ b/s/Hz, the decoding error probability for URLLC transmissions is $\epsilon_i = 10^{-5} ~\forall i \in \set{U}$, and the sensing SINR threshold is $\Gamma_{\text{th}}^{\text{sen}} = 6$ dB. The power consumption model parameters are in accordance with \cite{Chien2020Joint}: each antenna consumes $0.2$ W with a fixed fronthaul link power consumption of $0.825$ W for a total static-power consumption per AP of $P_{0} = 4.825$ W. The traffic-dependent fronthaul power is $P_{k}^{\text{tra}} = 0.25$ W/Gbps. The results are averaged across $100$ setups of uniformly random UE positions. We use the maximum ratio transmission (MRT) and regularized zero-forcing (RZF) precoders to show the efficacy of the proposed JPALB algorithm. We benchmark our results against a baseline problem that minimizes power consumption subject to URLLC rate requirements, sensing SINR constraints, and per-communication-AP power budget constraints, with \emph{no load balancing} (NLB). 

    In Fig. \ref{power-vs-sinr}, we plot the empirical cumulative distribution function (CDF) of the total power consumed (transmission power, static power consumption, and traffic-dependent fronthaul power consumption) in the network for the MRT and RZF precoding schemes, averaged across $100$ UE positions. In the case of NLB, the average power consumed is $57.95$ W for MRT and $58$ W for RZF. The proposed JAPLB algorithm with MRT and RZF requires an average power of $38.74$ W and
    $38.9$ W respectively leading to a $\sim33\%$ decrease in power consumption in comparison to their NLB counterparts. These average values are calculated as the expected values using the empirical CDF in Fig. \ref{power-vs-sinr}. The reason for the total power consumed in the network to go down is that the proposed JPALB algorithm switches off those communication APs that do not contribute much towards meeting the URLLC rate and sensing SINR requirements. In particular, the proposed JPALB algorithm turns off $\sim 37.5\%$ of the communication APs in the network when compared to the baseline NLB case, which leads to approximately $33\%$ reduction in the average total power consumed in the network when compared to the NLB case. Note that this power reduction happens without compromising the performance, specifically the constraints of the optimization problem in \eqref{opti_original_problem} ensure that the communication and sensing QoS requirements are always met. 

    The power consumed by the various components in the network is shown in Fig. \ref{bar-plots}. The major power consumed comprises the traffic-independent static power consumed by the fronthaul and the hardware power consumed by the antennas in the network. When the proposed JPALB turns off APs, the major power gains are thus achieved by saving on these two components.

    
    Fig. \ref{power-vs-rcs} shows the total power consumed in the network as a function of sensing $\sinr$ threshold $\left(\Gamma_{\text{th}}^{\text{sen}}\right)$, showing that the sensing requirements in the ISAC network are also satisfied. As $\Gamma_{\text{th}}^{\text{sen}}$ increases, the power consumed increases in order to satisfy the stronger sensing $\sinr$ constraint. This is because more number of communication APs need to be kept on to meet the increasing sensing $\sinr$, as this leads to more signals being reflected off the target and captured by the $R$ sensing APs (where $R$ is fixed). 

    \section{Conclusion}
    We designed an iterative joint power allocation and load balancing algorithm to minimize the total average power consumed in a cell-free massive MIMO network serving URLLC users and simultaneously detecting the presence of a target. For both MRT and RZF precoders, the proposed algorithm reduces power consumption in the network by turning off communication APs that do not contribute much towards meeting the URLLC rate and sensing SINR requirements. The proposed JPALB algorithm turned off $\sim37.5\%$ of communication APs, leading to a $33\%$ reduction in total power consumed by the network, when compared to a baseline case where the URLLC rate and sensing requirements are met without turning off any of the communication APs.
	
	\bibliographystyle{IEEEtran}
	\bibliography{document}
	
\end{document}